\documentclass{jfm}
\pdfoutput=1
\usepackage{graphicx}
\usepackage{epstopdf, epsfig}
\usepackage{amssymb,amsmath}
\usepackage{soul,color}
\usepackage{chngcntr,float}
\usepackage{float}

\shorttitle{Constrasts Between Momentum and Scalar Transport}
\shortauthor{Q. Li and E. Bou-Zeid}

\title{Contrasts Between Momentum and Scalar Transport Over Very Rough Surfaces}

\author{Qi Li\aff{1,2}\corresp{\email{ql56@cornell.edu}}
	\and Elie Bou-Zeid\aff{1}}

\affiliation{\aff{1}Department of Civil and Environmental Engineering, Princeton University,
Princeton, USA
\aff{2}School of Civil and Environmental Engineering, Cornell University,
Ithaca, USA
}

\begin{document}

\maketitle

\begin{abstract}
Large-eddy simulations are conducted to contrast momentum and passive scalar transport over large, three-dimensional roughness elements in a turbulent channel flow. Special attention is given to the dispersive fluxes, which are shown to be a significant fraction of the total fluxes within the roughness sublayers. Based on point-wise quadrant analysis, the turbulent components of the transport of momentum and scalars are found to be similar in general, albeit with increasing dissimilarity for roughnesses with low frontal blockage. However, strong dissimilarity is noted between the dispersive momentum and scalar fluxes, especially below the top of the roughness elements. In general, turbulence is found to transport momentum more efficiently than scalars, while the reverse applies for the dispersive contributions. The effects of varying surface geometries, measured by the frontal density, can be pronounced on both turbulent and dispersive fluxes. Increasing frontal density induces a general transition in the flow from a rough boundary-layer type to a mixed-layer-like type. This transition results in an increase in the efficiency of turbulent momentum transport, but the reverse occurs for scalars due to reduced contributions from large scale motions in the roughness sublayers. This study highlights the need for distinct parameterizations of the dispersive scalar fluxes and the importance of considering the contrasts between momentum and scalar transport for turbulent flows over very rough surfaces.
\end{abstract}

%\begin{keywords}
%	rough-wall turbulent boundary layers; dispersive flux; passive scalar, atmospheric flows, roughness sublayer
%	\hl{in JFM you should leave this blank and pick from the ones specified int he system during submission I think.}
%\end{keywords}

\section{Introduction}
The dynamics of turbulent shear flows over rough walls has been an active area of research because of its relevance in the design of engineering systems and in environmental fluid mechanics. Momentum and scalar exchanges between the wall and the fluid in such flows are of interest in a wide range of disciplines \citep{belcher2012wind}. Previous field experiments over natural vegetation \citep{poggi2004note,Katul:1997tkba}, wind tunnel studies over obstacles of regular shapes \citep{Macdonald1998,Castro:2006jbba}, and numerical simulations over three-dimensional roughness elements \citep{Coceal:2007fbba,Finnigan:2009ha,Leonardi:2015drba} have advanced our understanding of this problem significantly. For example, these past studies have underlined the importance of dispersive fluxes inside and close to the roughness elements \citep{poggi2008effect,poggi2004note,Jelly2018}. Nevertheless, there remains significant gaps in our knowledge, particularly concerning the transport of scalars and how it compares to that of momentum, over surfaces that consist of ``large" three-dimensional bluff-body-type roughness elements. Large here implies that the roughness protrudes significantly into the inertial layer and that the details of the flow below the top of the roughness elements, often called the canopy sublayer, are important for the application. In particular, \citet{Jimenez:2004dnbaca} in his review paper limited the discussions of rough-wall boundary layers to $H/\delta<0.025$, where $H$ is the roughness element height and $\delta$ is depth of the boundary layer. In many natural settings and engineering applications, $H/\delta$ often exceeds 0.1, a regime sometimes termed the \textit{very-rough} surface \citep{Castro:2006jbba}. 

Various previous numerical studies 
(e.g. \citet{Kanda:2004ftba,Castro:2006jbba,Coceal:2007fbba,Leonardi:2008hrba,Anderson:2015fvba,Giometto:2016exba,anderson2016amplitude,Li:2016cuba,Li:2016CHTC}) have probed the details of these very-rough-wall flows such as the morphology of coherent structures and the effects of the roughness in such regimes. Moreover, recent large eddy simulations (LES) and direct numerical simulations (DNS) \citep{Finnigan:2009ha,Boppana:2010tsba,Boppana:2012dfba,Philips:2013emba,Baik:2013tnba,Leonardi:2015drba,Li:2016CHTC} investigated the transport of scalars. However, compared to the extensive number of previous studies focusing on the flow and the momentum transport, research on scalars and their transport in flows over very rough walls remains quite limited. In addition, a detailed comparative analysis of momentum and scalar transport dynamics has not been performed before. These open gaps motivate this present paper: we investigate both momentum and scalars at very-high Reynolds numbers over three-dimensional, large roughness using the LES technique. The region within the roughness elements is defined as the canopy sublayer; the region just above the canopy where the flow is horizontally inhomogeneous is called the roughness sublayer. This paper focuses on these two regions. A logarithmic layer will exist further aloft if the roughness sublayer does not extend all the way to the top of the inertial layer. In particular, we focus on two aspects of the problem, namely \textit{effects of different roughness geometries} and \textit{comparisons between momentum and scalar transport}. After descriptions of the numerical setup of the problem in Section two, Section three analyzes the changes in turbulent flow characteristics for different surface geometries; the spatially-coherent dispersive fluxes are then investigated with a focus on the differences between momentum and scalar transport. Section four concludes with a summary and discussion.

\section{Numerical setup}
The LES code uses the immersed boundary method (IBM) to account for the presence of the roughness elements, in which a discrete-in-time momentum forcing is used to simulate the immersed boundary force \citep{Tseng:2006fgba,Chester:2007gbbacada,Li:2016cuba}. The non-dimensional filtered incompressible continuity (\ref{continuity}), Navier-Stokes (\ref{NS}), and scalar conservation (\ref{Scalareq}) equations are solved assuming hydrostatic equilibrium of the mean flow:
\begin{equation}\label{continuity}
\frac{\p u_i}{\p x_i}=0,
\end{equation}
\begin{equation}\label{NS}
\frac{\p u_i}{\p t}+u_j\left(\frac{\p u_i}{\p u_j}-\frac{\p u_j}{\p u_i}\right)=-\frac{\p p}{\p x_i}-\frac{\p \tau_{ij}}{\p x_j}+F_i+B_i,
\end{equation}
\begin{equation}\label{Scalareq}
\frac{\p \theta}{\p t}+u_i\frac{\p \theta}{\p x_i}=-\frac{\p q_i^s}{\p x_i}.
\end{equation}
All the variables we will discuss are filtered components, so the usual tilde above the symbols is omitted for simplicity. The density $\rho$ is taken to be unity and is uniform: buoyancy forces are not considered. $x$, $y$ and $z$ denote the streamwise, cross-stream and wall-normal directions respectively, and $u$, $v$ and $w$ are the velocity components in these respective directions. $t$ denotes time; $u_i$ is the resolved velocity vector; $x_i$ is the position vector; $p$ is a modified pressure \citep{Bou-Zeid:2005vaba}; $\tau_{ij}$ is the deviatoric part of the subgrid stress tensor; $B_i$ is the immersed boundary force representing the action of the obstacles on the fluid; and $F_i$ is the body force driving the flow (here simply a homogeneous steady horizontal pressure gradient along the $x$ direction with magnitude $u_*^2/\rho\delta$). A friction velocity of $u_*=1$ m s$^{-1}$, and a half channel width of $\delta$=100 m, are used to non-dimensionalize $t$ and all outputs of the code; their imposed numerical values are thus inconsequential here. The top boundary is impermeable with zero stress; the simulations therefore are similar to half a channel. In Eq. (\ref{Scalareq}), $\theta$ denotes a passive scalar quantity, which for illustration is considered to be temperature in Kelvin in the current simulation, and $q_i^s$ is the $i$th component of the subgrid scale scalar flux. Further numerical details on the code and the subgrid-scale model can be found in \citet{Bou-Zeid:2005vaba}, while detailed validations for the flow and scalar transport can be found in \citet{Li:2016cuba} and \citet{Li:2016CHTC}, respectively. The LES uses a wall-model for momentum and scalars that has been developed for a hydrodynamically smooth walls (here each facet of a building/cube is such a smooth wall) at high Reynolds numbers based on \citet{Yaglom:1972guba}. A constant surface temperature scalar boundary condition is used. Wall modelling for complex walls remains an on going area of research \citep{Yang:2015dzba} and an open challenge \citep{bose2018wall}; however, the performance of the current approach has been evaluated and shown to be quite satisfactory in \citet{Li:2016CHTC}.

\begin{figure}
	\centerline{\includegraphics[scale=0.6]{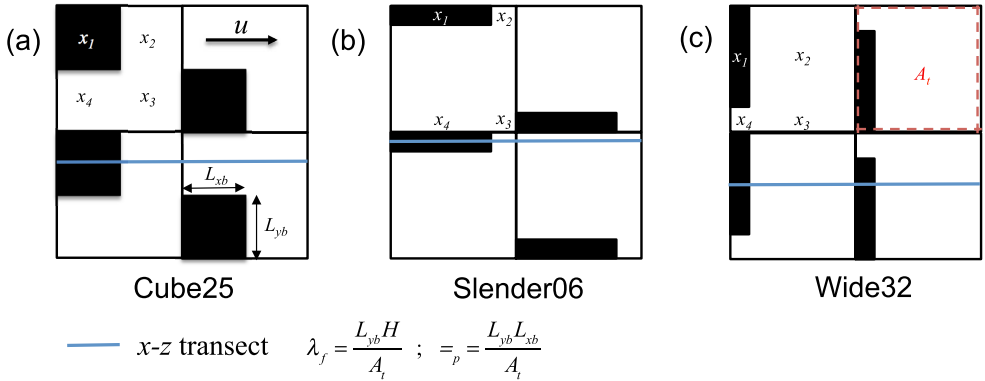}}% Images in 100% size
	\caption{Top-view of the `repeating unit' for three cases shown in table \ref{table1}, where shaded areas represent the obstacles: (a) Cube25; (b) Slender32; (c) Wide32. Area highlighted by red-dotted line is the lot area, $A_t$. Different intermediate cases labeled as Sf in table \ref{table1} are achieved through varying $L_{xb}$ and $L_{yb}$ while keeping $H$, the obstacle height, constant. Frontal area density, $\lambda_f$, and plan area density $\lambda_p$ are defined. Points labeled as $x_1$ to $x_4$ represent locations where instantaneous data are recorded for analysis.}
	\label{fig:setup}
\end{figure}

To explore the effects of the geometry, we conducted simulations of different cases as summarized in table \ref{table1} and illustrated in figure \ref{fig:setup}. $\lambda_f$ and $\lambda_p$ are the frontal area ratio and plan area ratio, respectively. $\lambda_f$ is defined as the total projected frontal (mean-flow normal) area of the roughness elements per unit wall-parallel area (i.e. land area); $\lambda_p$ is the ratio between the crest plan area (i.e. roof area, shaded in black figure \ref{fig:setup}) and the wall-parallel area. The first case is the classic cuboid obstacles. In the remaining cases, we gradually change the horizontal aspect ratio of the obstacles, while maintaining the same height and area ($\lambda_p$). Figure \ref{fig:setup} shows the cubic case and the two extreme aspect ratios of the other cases. All cases shown in Table 1 were run for 50 eddy turn-over times ($T\approx \delta/u_*$) and averaged for the last 25 $T$. The averaging time is comparable to that in \citet{leonardi2010channel}, where averaging time is 2000$T_b$ for $T_b=H/U_b$; $U_b$ is the bulk streamwise velocity and $H$ is the obstacle height. 
Throughout the paper, $\left\langle X\right\rangle$ defines a volume average of $X$ over $L_xL_y\Delta z$, for $\Delta z$ being the vertical grid size; $\overline{X}$ defines temporal average of $X$. 

\begin{table}
	\begin{center}
		\def~{\hphantom{0}}
		\begin{tabular}{lcccccc}
			Case   & $\lambda_p$ & $\lambda_f$ & $N_x^b,N_y^b,N_z^b$   & $N_x,N_y,N_z$ & $L_x/\delta$ & $L_y/\delta$ \\[3pt]
			\hline
			Cube25   &0.25 &0.25 & ~8,~8,~8 & 192,~96,~64 & 3 & 1.5\\
			Slender06 &0.12 &0.06  & 16,~3,~8 & 200,~100,~64 & 3.125 & 1.5625\\
			Sf08     &0.12 &0.08 & 12,~4,~8 & 200,~100,~64 & 3.125 & 1.5625\\
			Sf12     &0.12 &0.12 & 8,~6,~8 & 200,~100,~64 & 3.125 & 1.5625\\
			Sf16     &0.12 &0.16 & 6,~8,~8 & 200,~100,~64 & 3.125 & 1.5625\\
			Sf24     &0.12 &0.24 & 4,~12,~8 & 200,~100,~64 & 3.125 & 1.5625\\
			Wide32    &0.12 &0.32 & ~3,~16,~8 & 200,~100,~64 & 3.125 & 1.5625\\
		\end{tabular}
		\caption{Summary of simulation parameters. $N_i^b$ is the number of grid points resolving one obstacle in the $i$ direction, while $N_i$ is the total number of computational points in that direction. $L_x$ and $L_y$ are the domain dimensions in the streamwise and cross-stream directions. The two digits at the end of each case name are $\lambda_f\times100$ for that case. Sf stands for `Stagger frontal'.}
		\label{table1}
	\end{center}
\end{table}

A sensitivity test was carried out to examine the effect of the number of points used to represent the obstacles and the domain length, particularly for cases where one dimension of the cubes is only spanned by three grid points. The comparisons for case Wide32 between the original simulation, a doubled resolution run and a run with doubled $L_x$ are shown in the appendix. Both simulations show converged results in the canopy sublayer and roughness sublayer. We thus conclude that the numerical resolution is sufficient \it{for the present purposes}\rm.

\section{Results and discussion}
\subsection{Turbulent transport}
\subsubsection{Effects of surface geometry}
\begin{figure}
	\centerline{\includegraphics[scale=0.6]{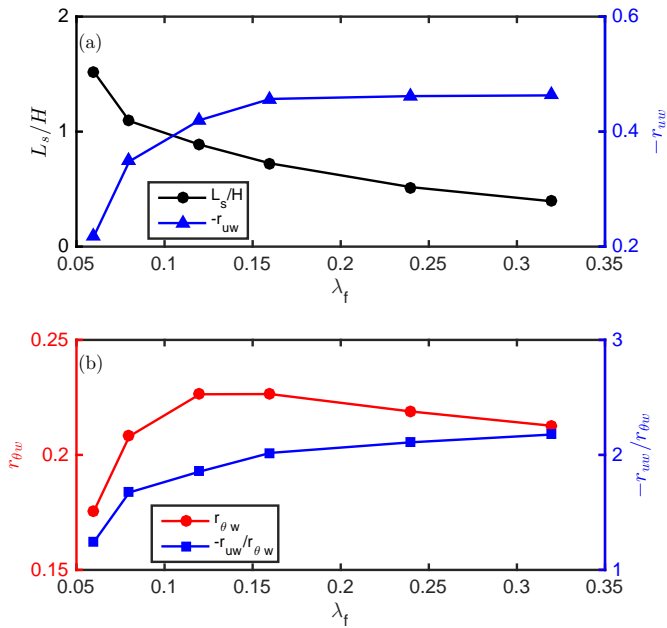}}% Images in 100% size
	\caption{(a): Shear length scale, $L_s$ and the correlation coefficient $r_{uw}$ for cases listed in table \ref{table1} (except cube25) computed at $z=H$; (b) $r_{\theta w}$ and $r_{uw}/r_{\theta w}$  for the same cases as (a).}
	\label{fig:Lsruw}
\end{figure}

We first focus on the impact of surface geometry on the general turbulent flow characteristics. The parameter space that characterizes the surface geometry is large. A non-exhaustive list of studies includes previous works that have investigated height variations \citep{Yang:0wuba}, geometric shapes \citep{Leonardi:2015drba,Yang:0wuba,Llaguno-Munitxa2017,Llaguno-Munitxa2018} of regular \citep{Kanda:2004ftba,Ganapathisubramani:2015bxba} or irregular \citep{Chester:2007gbbacada,yuan2018topographical} surface roughness elements, as well as statistical moments of roughness elements \citep{Zhu2017}. We do not aim to comprehensively examine the parameter space in this paper, but we are more interested in the general transition of the flow as the roughness, conserving the same area density $\lambda_p$, changes from slender elements with low $\lambda_f$ to wide ones with high $\lambda_f$. \citet{GHISALBERTI:2009glba} analyzed the dynamics of flows over many different types of canopies and used the term `obstructed shear flow' to characterize them. It is found that this type of canopy flow can be characterized by the penetration depth ($\delta_e$) of the vortices into the canopy sublayer. Figure \ref{fig:Lsruw}a shows the shear scale $L_s$ as a function of the frontal blockage ratio $\lambda_f$. $L_s$ is a basic length scale in canopy flows similar to the vorticity thickness defined in a plane mixing layer; it is defined as $L_s=\frac{\langle\bar{u}\rangle}{\langle\bar{u}/dz\rangle}$, here computed at $z=H$. $L_s$, which is analogous to the penetration depth $\delta_e$, decreases with increasing $\lambda_f$ (figure \ref{fig:Lsruw}a), which signifies a larger shear strength and hence more deviation from the classical surface layer profile as the blockage increases. $L_s$ for Slender 06 and Wide32 differ significantly and additional intermediate cases show a monotonic decrease with $\lambda_f$.

The correlation coefficient $r_{uw}$ computed as $\left\langle\frac{\overline{u^{\prime}w^{\prime}}}{\sigma_{u}+\sigma_{w}}\right\rangle$ at $z=H$ is also depicted in figure \ref{fig:Lsruw}a. Its magnitude can be interpreted as a vertical turbulent momentum transport efficiency \citep{Bou-Zeid:2011hrba}. As $\lambda_f$  increases, $-r_{uw}$ increases approximately from 0.2 to 0.46, the latter being consistent with what is typically found in canopy flow and mixing layer flow ($-r_{uw}\approx0.5$) \citep{Finnigan:2003igbaca}.  It is worth noting that obstacles in all the cases presented in figure \ref{fig:Lsruw} have the same plan area density (i.e. $\lambda_p=0.12$); but by changing $\lambda_f$, we are observing a transition from a canonical boundary-layer flow to a mixing-layer one where momentum turbulent transport is more efficient. As this transition occurs, and unlike for its momentum counterpart, the vertical transport efficiency for scalar exhibits a non-monotonic behaviour (figure \ref{fig:Lsruw}b), reaching a peak of approximately 0.23 at around $\lambda_f=0.12$. For large $\lambda_f$, while momentum $r_{uw}$ is more than twice larger than its value at $\lambda_f=0.06$, the scalar $r_{\theta w}$ changes less significantly with changing geometry. Overall, the results show that as the flow becomes more mixing-layer like, the momentum and scalar transport efficiencies increasingly diverge. The ratio of their correlation coefficients become appreciably larger than 1 (figure \ref{fig:Lsruw}b), making the Reynolds analogy (which postulates that momentum and scalars are transported similarly) less applicable. Note that we do not imply that the very rough canopy or vegetation canopy completely resembles the canonical mixing layer flows. Rather, the examples shown in the present study demonstrate that the rough wall-bounded flows can exhibit mixed properties of wall-bounded and mixing-layer flows \citep{Kanda2006}.

\begin{figure}
	\centerline{\includegraphics[scale=0.6]{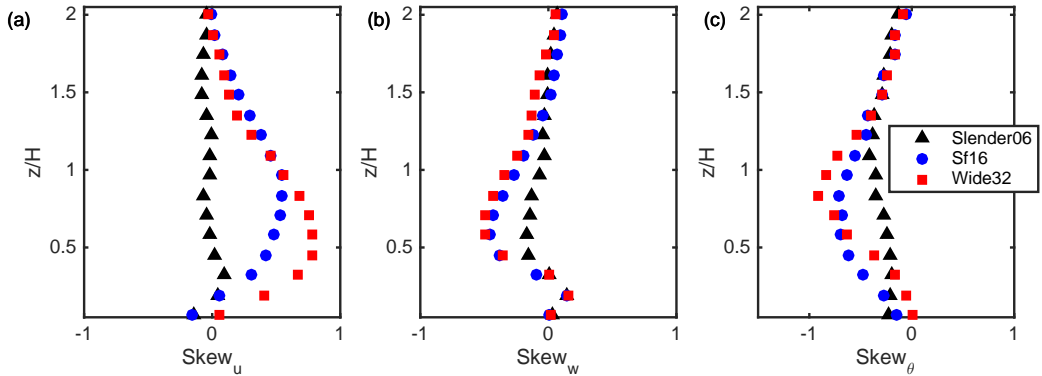}}% Images in 100% size
	\caption{Skewness of fluctuating quantities at $x_3$: (a) $u^\prime$; (b) $w^\prime$; (c) $\theta^\prime$
	}
	\label{fig:skew}
\end{figure}

To gain a better understanding of the turbulent statistics at representative points with respect to the roughness elements, we sampled data at four different points  $x_1$-$x_4$ (see figure \ref{fig:setup}) for all heights. The point-wise time series of relevant quantities are recorded at a frequency of $1/(2500 T)$, where $T$ is the eddy turn-over time defined previously. Data were sampled at each time step for a total period of 2$T$. This time averaging alone is not sufficiently long to guarantee complete statistical convergence of the results. However, the statistics were also averaged for all repeating units to improve statistical convergence and the primary characteristics of the flow that we will focus on are already very clear with this limited averaging time. We first examine the skewness of the fluctuating components shown in figure \ref{fig:skew} for point $x_3$ only for $u/u_*$, $w/u_*$ and $\theta/\theta_1$, where $\theta_1=\theta(z=\delta)$. We only illustrate three cases with the same $\lambda_p$, spanning the full range of $\lambda_f$. Slender06 has close to zero skewness for $u^\prime$ and $w^\prime$ (figures \ref{fig:skew}a and b) in the roughness sublayer but higher values are observed as $\lambda_f$ increases in cases Sf16 and Wide32, more typical of canopy flows from experimental measurements over vegetation canopies \citep{Raupach:1981etba,Rotach:1993kkba,Finnigan:2003igbaca}. The variation in skewness presents further evidence of the role of frontal area, and hence pressure drag, in dictating the characteristics of the roughness sublayer dynamics. The negative skewness for $\theta^\prime$ in figure \ref{fig:skew}c signifies dominance of downward events that tend to bring down cooler fluid regardless of the underlying surface geometry (this is also consistent with the negative skewness of $w^\prime$ in figure \ref{fig:skew}b). This is similar to findings for temperature, as a passive scalar, over rough surfaces in the DNS by \citet{Leonardi:2008hrba}.

Since in physical space we observed that changes in surface roughness modify the characteristic length scale and turbulence statistics, we now examine in spectral space if that also implies a modification in the range of length scales that contributes the most to the turbulent transport. In spectral space, the correlation spectrum is defined as $R_{XY}=Co_{XY}/\left( \Gamma_X\Gamma_Y\right)$, where $Co_{XY}$ is the cospectrum of time series $X$ and $Y$ and $\Gamma_X$ and $\Gamma_Y$ are the power spectra of $X$ and $Y$, respectively.
\begin{figure}
	\centerline{\includegraphics[scale=0.6]{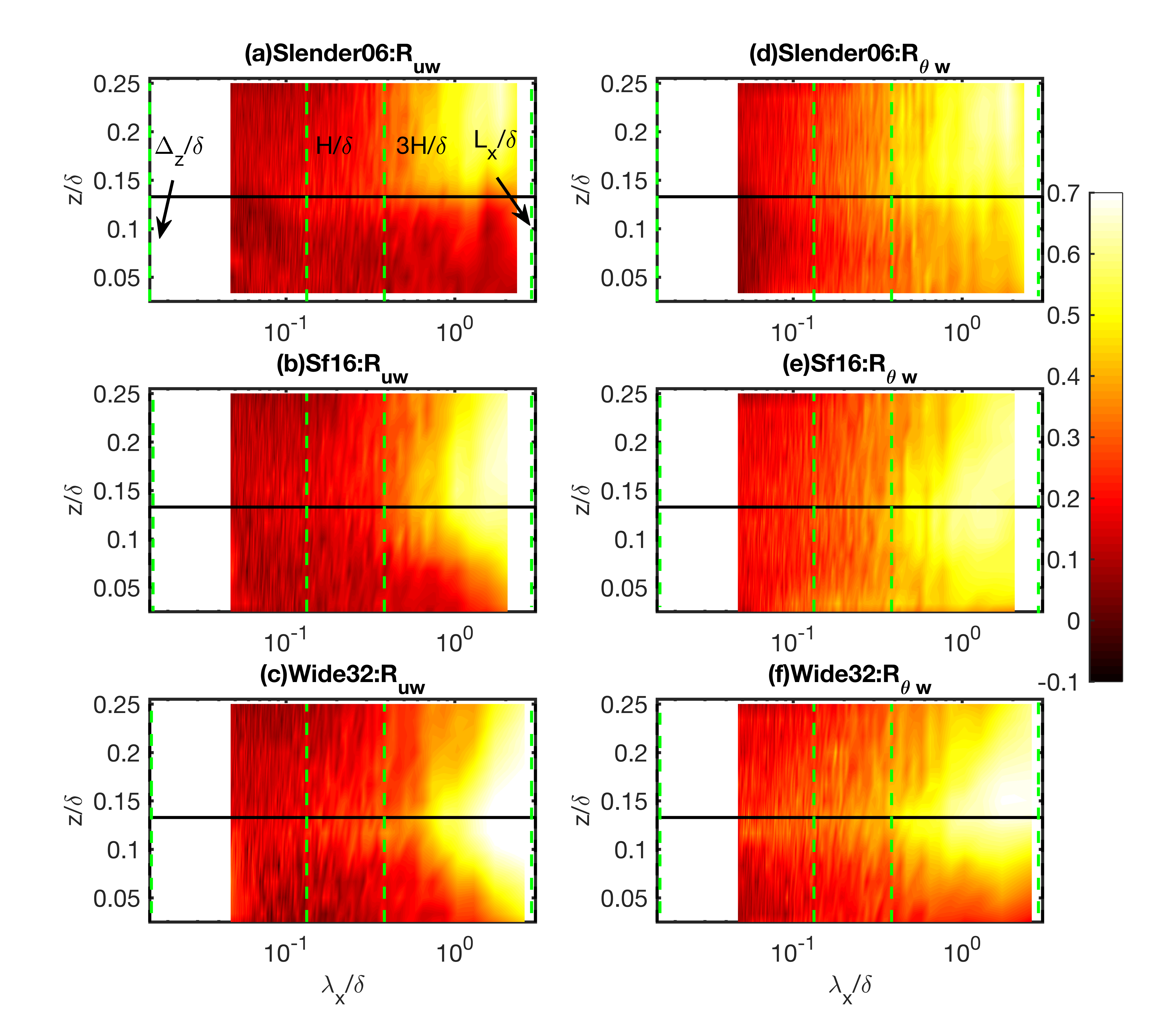}} %{Fig8}}% Images in 100% size
	\caption{Pseudocolour plot of the correlation spectra $R_{uw}$ in (a)-(c) and $R_{\theta w}$ in (d)-(f) for different cases $x_3$. The dotted vertical green lines correspond to $\Delta_z/\delta$, $H/\delta$, $3H/\delta$ and $L_x/\delta$ as labeled in (a); solid horizontal black line denotes the top of the obstacles (illustrated in (a)).
	}
	\label{fig:RXY}
\end{figure}
For $Y$ being vertical velocity fluctuation $w^\prime$ and $X$ being $u^\prime$ or $\theta^\prime$, $R_{XY}$ is interpreted as a wavelength specific transport efficiency in the spectral space. The total length of the time series defined in time unit $L_z/U_0$ is about 60, where $U_0$ is the free stream velocity. Using Taylor's frozen turbulence hypothesis to convert time to length scale with $U_0$, and performing ensemble averaging over 20 shorter time series (obtained by sectioning the total time series into 20 sub-series) at each representative point $x_i$, the correlation spectra are computed. The results are shown in figure \ref{fig:RXY}, where only results at $x_3$ (in front of the obstacle) are presented; similar conclusions can be obtained from the results at other points (not shown here). 
Eddies of large scales (close to $\delta=L_z$) contribute to the efficient transport of both momentum, and scalars, but a wider range of scales contribute to $R_{\theta w}$. Higher correlation spectra values can be noted for scalars, compared to momentum, for all three cases between $H/\delta$ and $3H/\delta$. As the roughness frontal blockage increases (e.g. for case Wide32), the large scale motions ($O(\lambda_x/\delta)=1$) can penetrate into the canopy more easily and they become more efficient in transporting momentum below the canopy top (figures \ref{fig:RXY}a to c). For scalars, however, it seems they are most efficient at transport deep inside the canopy at intermediate densities $\lambda_f$ (case Sf16), becoming less efficient in case Wide32. 

In summary, figure \ref{fig:RXY} shows the change in length scales of the most-efficient transport eddies across different geometries. As the transition to a mixed-layer-type flow indicated in figure \ref{fig:Lsruw} occurs, larger scale motions contribute more effectively to momentum transport in the canopy and roughness sublayer. However, for scalar transport, there seems to be an optimal configuration for the large scalar motions to interact effectively with the lower part of the canopy.\\

\subsubsection{Flow and passive scalar structures}

\begin{figure}
	\centerline{\includegraphics[scale=0.6]{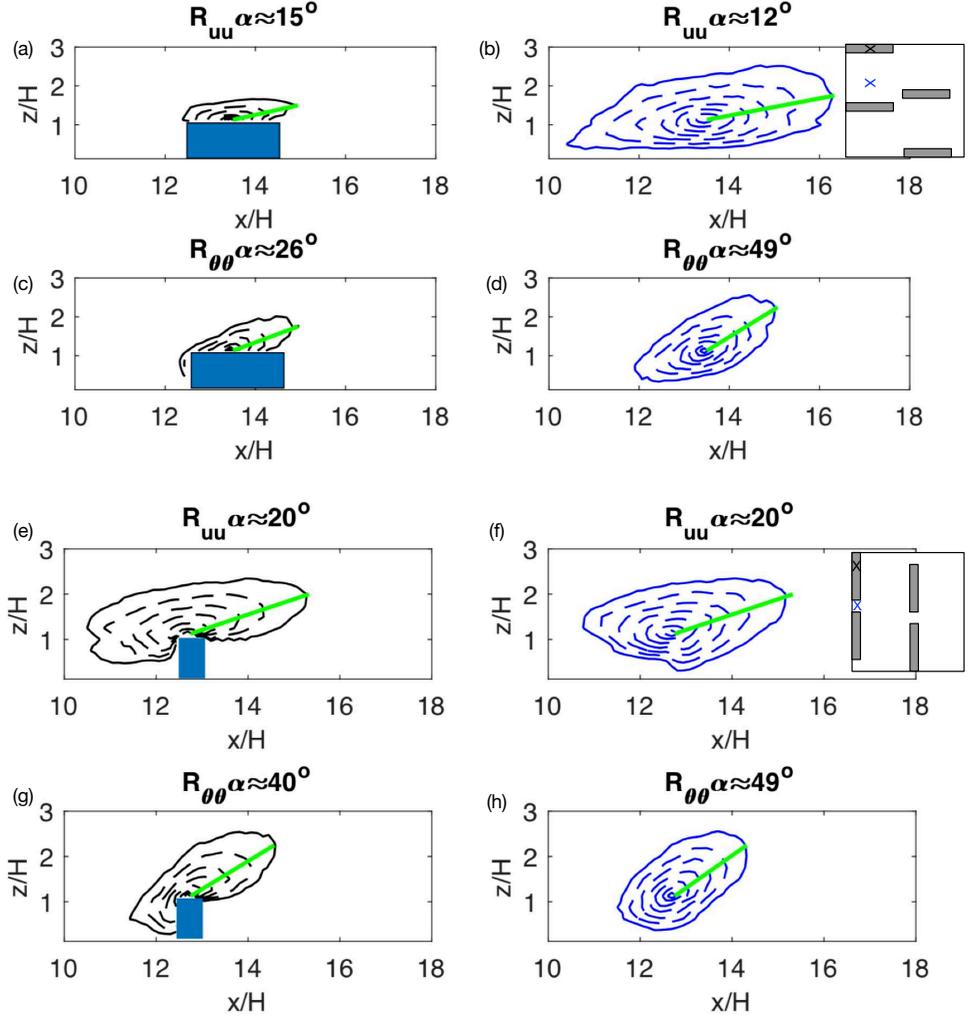}}% Images in 100% size
	\caption{Two-point streamwise correlation for $u$ and $\theta$ for case Slender06: (a)-(d) and case Wide32: (e)-(h). The black (blue) crosses in the inset figures indicate the reference points $X_0$ in figures a, c, e and g (b, d, f and h). The correlation contours are spaced uniformly at 0.1 from 1 to 0.3, where the outermost solid contour corresponds to $\rho_{ss}$=0.3. The solid green line segments are drawn between $X_0$ and $X_1$, where $X_1$ is on the correlation contour of 0.3. The angle $\alpha$ is between the streamwise direction and the line segment $X_0X_1$.}
	\label{fig:Ruuall}
\end{figure}

To understand how turbulent momentum and scalar transport behave as the flow regime transition illustrated in the previous sub-section occurs, we analyze the spatial correlations of turbulent quantities. The two-point streamwise correlation for a quantify $s$ at reference point $x$ and for a separation distance $X_0$ can be computed as $\overline{\rho_{ss}}\big(X_0,x\big) = \overline{s^\prime\big(X_0\big)s^\prime\big(x\big)}/\big(s_{rms}(X_0)s_{rms}(x)\big)$, where rms denotes the root mean square of the turbulent fluctuations. These correlations are shown in figure \ref{fig:Ruuall} for cases Slender06 and Wide32. The angle of inclination of the contours was calculated following the approach of \citet{Leonardi:2015drba}, where $\alpha$ is estimated as the angle between the horizontal direction and the segment $X_0X_1$, indicated by the green solid line, the point $X_1$ being the furthest from $X_0$ on the contour $R_{uu}$ =0.3. The increased interactions between the canopy top and the roughness sublayer are indicated by larger $\alpha$, especially $R_{uu}(x_2)$ in figure \ref{fig:Ruuall}f compared to figure \ref{fig:Ruuall}b.  Compared to the two-dimensional bars studied in \citet{Leonardi:2015drba}, the three-dimensional roughness geometry and different configurations also introduce spanwise variation in the correlation contours. The mixing-layer like flow regime results in increased interactions between the surface and the roughness sublayer for both $u^\prime$ and $\theta^\prime$. However, the increase is more pronounced for the velocity, indicating that the correlation contours $R_{\theta\theta}$ are less sensitive to the increase of $\lambda_f$ than their velocity counterparts. This implies that even for a surface-layer like flow regime, where the coupling between canopy sublayer top and the roughness sublayer aloft is not as strong as that for the mixing-layer like case, the effect of the roughness geometry on passive scalar can extend further up. This result is also consistent with the larger scale of the motions contributing to $R_{\theta w}$  than to $R_{uw}$ in figure \ref{fig:RXY}.

\begin{figure}
	\centerline{\includegraphics[scale=0.45]{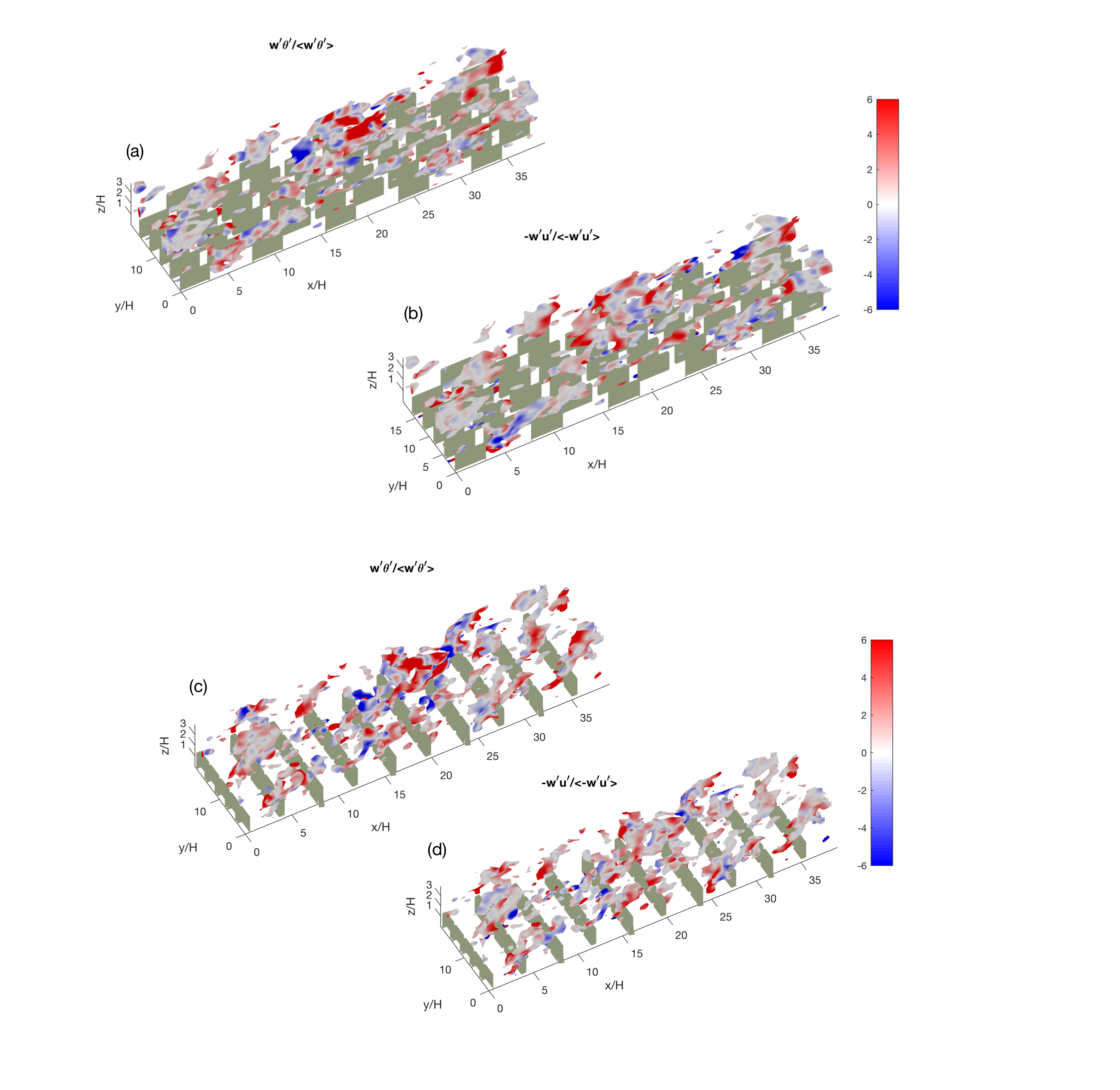}}% Images in 100% size
	\caption{Isosurfaces of $\theta^\prime_{high}$, where $\theta^\prime_{high}= \langle\theta^\prime(z=H)\rangle+1.5 Std(\theta^\prime(z=H)))$ low-pass filtered for better visualisation, for Slender06 (a) and (b) and Wide32 (d) and (d). $Std(\theta^\prime(z=H))$ denotes the standard deviation of a snapshot of $\theta^\prime$ at $z=H$.}
	\label{fig:fluxvisz}
\end{figure}

The iso-surfaces of high temperature fluctuation, $\theta^\prime_{high}$, are shown in figure \ref{fig:fluxvisz}. $\theta^\prime_{high}$ is defined here as $1.5 \theta^\prime_{rms}$ at any instant. However, here we colour these iso-surfaces by the normalized Reynolds stress and wall-normal scalar flux (figures \ref{fig:fluxvisz}b, d and figures \ref{fig:fluxvisz} e, f, respectively). First, although both Slender06 and Wide32 demonstrate streamwise thermal `streaks' \citep{Hetsroni1999,Leonardi:2015drba}, Wide32 is characterized by more patchy structures in most of the domain; whereas the temperature iso-surfaces in Slender06 are more elongated in the streamwise direction and are aligned in between the staggered roughness elements. Stronger flow disturbances by the obstacles are observed in the mixing-layer-like flow where the more obstructive three-dimensional geometry is found to significantly disrupt the structures, in agreement with the experimental observations in \citet{Hetsroni1999}. Second, there is a general spatial correspondence between $w^\prime\theta^\prime$ and $-w^\prime u^\prime$ ``events", especially in the roughness sublayers (see figures  \ref{fig:fluxvisz}a and b, $x/H$=10-15; figures  \ref{fig:fluxvisz}c and d). Nevertheless, for instantaneous flow structures, there can also be discernible dissimilarities between them in the canopy sublayer, suggesting that some eddies might be carrying momentum but not scalars or vice-versa. Thus, we conduct quadrant analysis in next section to further investigate the (dis)similarity of momentum and scalar.

\subsubsection{Quadrant analysis of turbulent momentum and scalar transport}
Quadrant analysis is a useful and widely used technique for probing how turbulent motions evolve and transport momentum and scalars in the wall-normal direction \citep{Wallace:2016gpba}. Thus, we apply this technique here to compare the momentum and scalar turbulent transports over the canopy and roughness sublayers. The definition of each quadrant for momentum flux follows previous studies \citep{Katul:1997veba,Katul:1997tkba,Bou-Zeid:2011hrba}. Q1 events are classified as $ s^\prime >0$ and $w^\prime > 0$; Q2 as $ s^\prime<0$ and $w^\prime > 0$; Q3 as $ s^\prime <0$ and $w^\prime < 0$; Q4 as $ s^\prime > 0$ and $w^\prime < 0$, where $s$ is $u$ or $\theta$. Here a prime denotes the turbulent perturbation of an instantaneous value from its Reynolds (time) average denoted by an overbar. We applied quadrant analysis to time series collected at four representative horizontal locations ($x_1$ to $x_4$) indicated in figure \ref{fig:setup} and at every height, but we will only show results for $x_2$ and $x_3$ since the other locations convey the same information, which are shown in the appendix. The same averaging procedure detailed for the results of the skewness is used here. For the momentum flux, Q2 and Q4 are termed ejections and sweeps respectively; for scalar flux, Q1 (motion that transports higher concentrations of the scalar upward, the wall being a source here) and Q3 (motion that brings lower concentrations downward) are termed ejections and sweeps, respectively. Compared to the other two quadrants, the ejections and sweeps are the dominant events in transporting momentum and scalars; they contribute to down-gradient transport.
\begin{figure}
	\centerline{\includegraphics[scale=0.72]{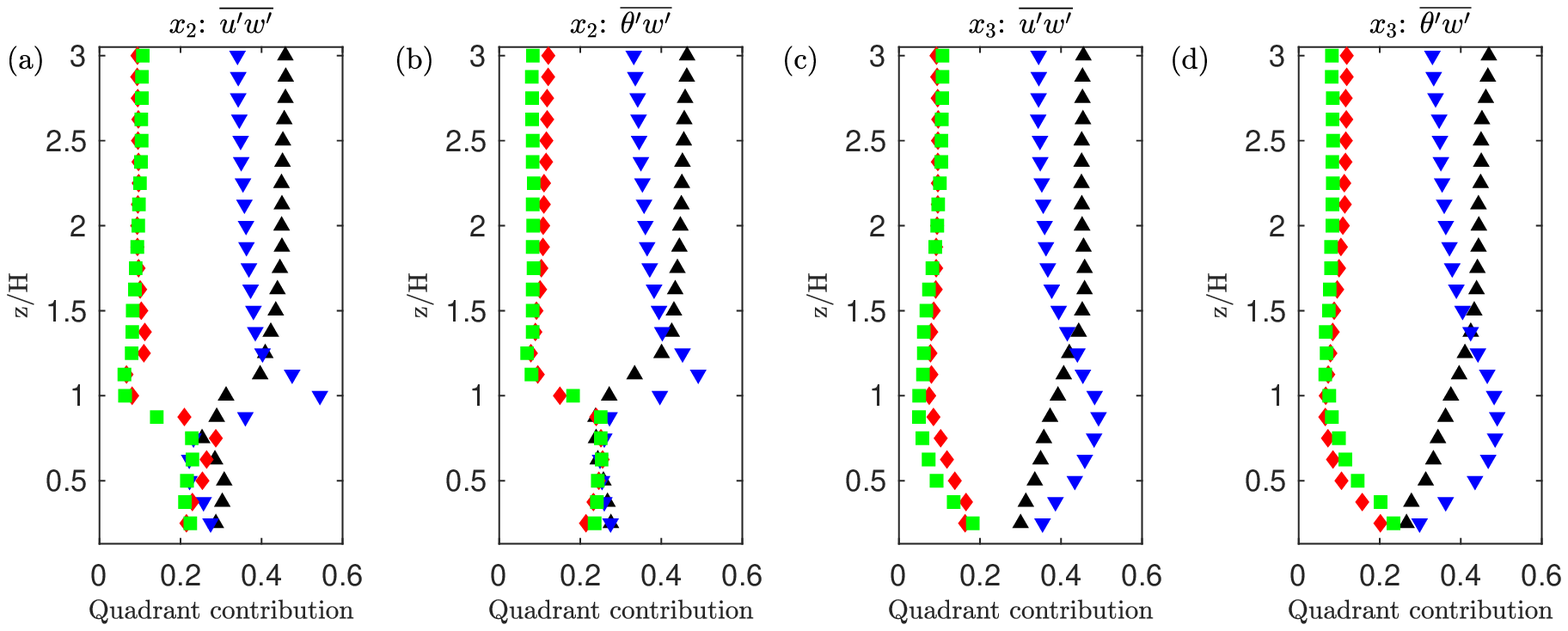}}% Images in 100% size
	\caption{Contribution to total flux for (a),(c) momentum and (b),(d) scalar at points $x_2$ (back) and $x_3$ (front) in case Cube25. 
		Ejections are black $\bigtriangleup$
		; sweeps are blue $\bigtriangledown$; outward interactions ($w^\prime>$0 and $u^\prime>0$ or $\theta^\prime<0$) are red $\diamond$; and 
		inward interactions ($w^\prime<$0 and $u^\prime<0$ or $\theta^\prime>0$) are green $\square$.}
	\label{fig:qcontribution}
\end{figure}

Figure \ref{fig:qcontribution} shows the contributions from various quadrants (momentum in (a) and (c); scalar in (b) and (d)) for case Cube25, where the contribution of quadrant $i$ is defined as $\lvert \overline{s^\prime w^\prime}\rvert_i /\ \sum_i\lvert \overline{s^\prime w^\prime}\rvert $. The contribution of each quadrant to scalar fluxes is broadly similar to that of momentum across all points, though slight differences between momentum and scalars are seen inside the canopy sublayer below $z/H=1$. The spatial variation across the four different horizontal location in the canopy sublayer is on the other hand significant, underlining the complexity of the flow within these three dimensional roughness arrays. These differences blend away above the canopy. The number of ejection events exceeds that of sweeps below $z/H=1.25$ across all four points (not all shown here), indicating that below this height sweeping events are stronger compared to the more frequent and less intense ejections. The crossover point at $z/H=1.25$ is consistent with results for the Reynolds stress from DNS reported by \citet{Coceal:2007fbba} for the same underlying roughness geometry of staggered cubes, although they averaged over all points at a given $z$. It has been previously observed that for momentum transport, the sweeps' contribution to total stress dominates over ejections, under near-neutral static stability and in the roughness sublayers, in vegetation canopies \citep{Patton:2012hpba} and in a realistic urban area characterized by large roughness elements \citep{Rotach:1993kkba,christen2007coherent}. Dominance of sweeps in momentum transport over very rough walls, compared to the dominance of ejections over smooth and less-rough counterparts, is in general agreement with the picture proposed and discussed in detail by \citet{Raupach:1996cjba} who adopt a mixing-layer analogy for these very-rough walls to account for the differences compared to a typical smooth-wall surface layer. In the outer layer, the dominance of ejections is restored as typical in smooth-wall boundary layers \citep{Adrian:2007hyba}. 

Our results for case Cube25 demonstrate that the turbulent transport of a passive scalar exhibits similar behaviour as for momentum. Similarity or dissimilarity between momentum and passive scalar transport over wall-bounded turbulent flows has also been discussed in the literature. For instance, similarity in the transport is found in laboratory results \citep{perry1976experimental,nagano1988statistical} over smooth surfaces and field measurements in the atmospheric boundary layer over vegetation canopy and in the roughness sublayer over an urban area  \citep{Patton:2012hpba,Wang:2014ksba}, and in DNS over two-dimensional square-shaped obstacles \citep{Leonardi:2015drba}; whereas dissimilarity is not as extensively reported \citep{christen2007coherent}. The generalizability of the similarity between turbulent transport of momentum and scalar observed above for case Cube25 to other cases is assessed through quadrant-analysis based turbulent transport efficiencies, defined as $\eta=F_{total}/\left(F_{ejection}+F_{sweep}\right)$ \citep{Moeng:1992gkba,Bou-Zeid:2011hrba}, where $F$ is the (total or from a single quadrant) flux of momentum or scalars. The results for cases Cube25, Slender06 and Wide32 are shown in figures \ref{fig:efficiency}-\ref{fig:efficiency2}. Figures \ref{fig:efficiency}-\ref{fig:efficiency2}(a) and (b) show the vertical profiles of efficiencies at each horizontal location, and an average over the four points $x_1$ to $x_4$, while figures \ref{fig:efficiency}-\ref{fig:efficiency2}(c) shows the ratio of the $\eta_m$ to $\eta_s$, averaged over the four horizontal locations. 

Above $z/H = 1.25$, $\eta_m-\eta_s$ approaches 0 and spatial variability is blended out at all points and across all cases shown in figures \ref{fig:efficiency}-\ref{fig:efficiency2}. Below that blending height, case Cube25 (figure \ref{fig:efficiency}) shows the most significant spatial variation of efficiencies across four points: $\eta_m$ and $\eta_s$ for Slender06 (figure \ref{fig:efficiency1}) vary much less while for Wide32 only $\eta$ at $x_4$ deviates appreciably from the other points. Although there are some discrepancies between $\eta_s$ and $\eta_m$ at some locations, such as at $x_2$ and $x_4$ within the canopy sublayer of Cube25 and Wide32, the general trends for momentum and scalar efficiency remain similar for those cases. On the other hand, Slender06 displays more dissimilarity between momentum and scalar in the canopy sublayer as shown by both $\eta$ values (figure \ref{fig:efficiency1}), with a higher transport efficiency for the scalar. While sweeps dominate the transport of the scalar in Slender06 (figure \ref{fig:quadrant_allpoints_scalar}), both sweeps and ejections contribute about equally to the total momentum flux (figure {\ref{fig:quadrant_allpoints_momentum}), especially at points $x_3$ and $x_4$. A physical explanation for this dissimilarity is that the penetration of high horizontal momentum sweeps downward is hindered by the blockage effect of the slender-shaped roughness element, which induces negative fluctuations in $u^\prime$ and thus increased inward interactions and reduced $\eta_m$. Nevertheless, sweeps that are associated with cold (low $\theta^\prime$) fluid but that do not have a large positive $u^\prime$ can still efficiently penetrate the canopy and transport the passive scalar. This Slender06 case is characterized by higher $\eta_s$ than $\eta_m$. Overall, turbulent momentum and scalar transport are broadly similar for Cube 25 and Wide 32, but not for Slender06. It would be instructive to analyze further why the slender blocks induce more dissimilarity than the wider ones, but this is not at the core of the aim of this paper and cannot be pursued here. What we can conclude is that, as the general characteristics of turbulent flows transitions from surface-layer-like to mixing-layer-like regimes, $u^\prime$ and $\theta^\prime$ exhibit more similar correlations with $w^\prime$. For these mixing-layer-like flows, turbulent momentum and scalar transports are broadly similar but with a slightly more efficient momentum transport ($\eta_m>\eta_s$). However, this similarity is reduced for surface-layer-like turbulence similar and the more efficient exchange is now for scalars with $\eta_m<\eta_s$). For all geometries, the results suggest a turbulent Schmidt or Prandtl number $\neq1$.

\begin{figure}
	\centerline{\includegraphics[scale=0.8]{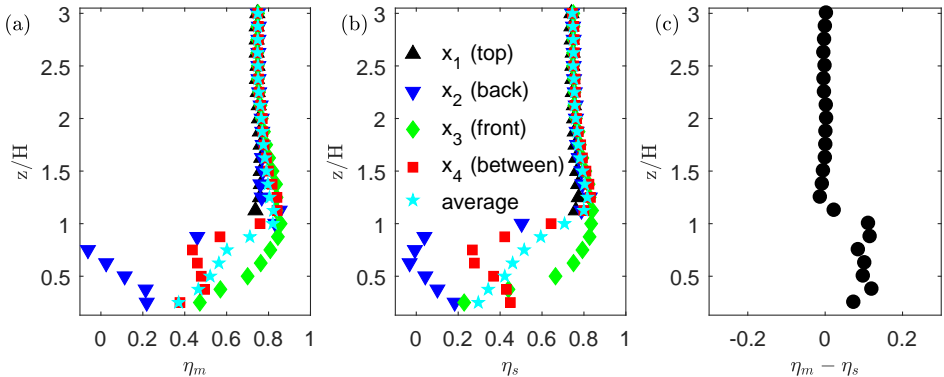}}% Images in 100% size
	\caption{Turbulent transport efficiencies at $x_1$-$x_4$ for the case of staggered cube (a) momentum transport efficiency $\eta_m$; (b) scalar transport efficiency $\eta_s$; (c) $\eta_m-\eta_s$ averaged over points $x_1$-$x_4$.}
	\label{fig:efficiency}
\end{figure}
\begin{figure}
	\centerline{\includegraphics[scale=0.8]{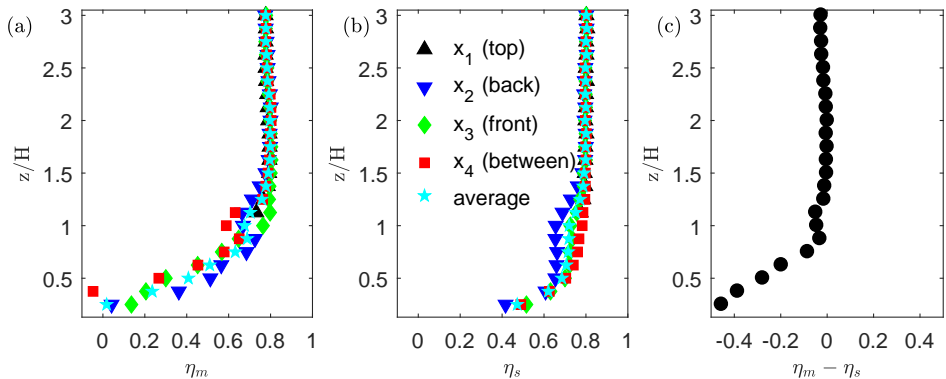}}% Images in 100% size
	\caption{Same as figure \ref{fig:efficiency}, except for case Slender06.}
	\label{fig:efficiency1}
\end{figure}
\begin{figure}
	\centerline{\includegraphics[scale=0.8]{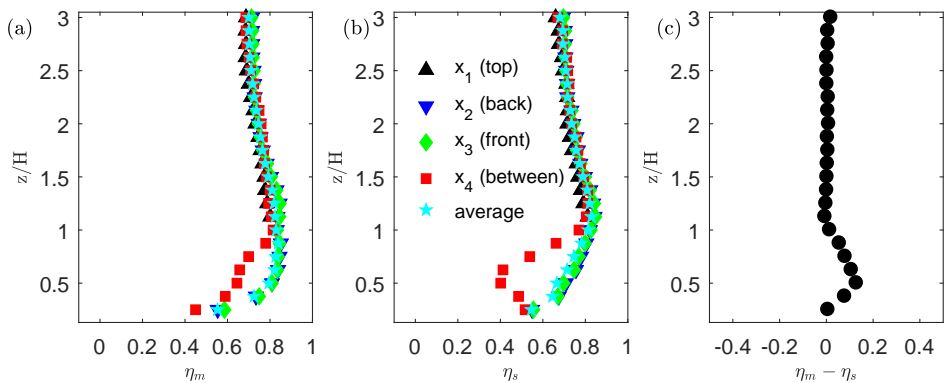}}% Images in 100% size
	\caption{Same as figure \ref{fig:efficiency}, except for case Wide32.}
	\label{fig:efficiency2}
\end{figure}

\subsection{Dispersive transport}
An important aspect that sets very rough walls apart is that dispersive stresses or fluxes, which arise from the spatial inhomogeneity of the time-averaged flow field, can be important contributors to total transport. In the multiply connected space inside the roughness arrays, the spatial averaging and differentiation operations do not commute \citep{Finnigan:1985kqba,Finnigan:2003igbaca}. Any mean (time-averaged) quantity $\phi$ can be decomposed into $\phi=\left\langle\phi\right\rangle + \phi^{\prime\prime}$, where the double prime represents the spatial deviation of the time-averaged variable from that of the spatial average. In the present paper, we consider $\left\langle\phi\right\rangle$ representing the planar average over an $x-y$ plane at a given height. The dispersive fluxes then arise from the spatial averaging of Eqs. (\ref{NS}) and (\ref{Scalareq}). Therefore, the spatially-local, time averaged dispersive momentum stress is $\overline{u_i^{\prime\prime}u_j^{\prime\prime}}$ and the dispersive scalar flux is $ \overline{u_i^{\prime\prime}\theta^{\prime\prime}}$, which can then also be spatially averaged over the plane.

Figure \ref{fig:disturb_ratio} shows the ratios 
$F_{dis}/F_{total}=\left\langle\overline{u^{\prime\prime}w^{\prime\prime}}\right\rangle /\left(\left\langle \overline{u^{\prime\prime}w^{\prime\prime}}\right\rangle + \left\langle \overline{u^{\prime}w^{\prime}}\right\rangle\right) $ and $F_{turb}/F_{total}=\left\langle\overline{u^{\prime}w^{\prime}}\right\rangle /\left(\left\langle \overline{u^{\prime\prime}w^{\prime\prime}}\right\rangle + \left\langle \overline{u^{\prime}w^{\prime}}\right\rangle\right)$  in (a) and their counterparts
$\left\langle\overline{w^{\prime\prime}\theta^{\prime\prime}}\right\rangle /\left(\left\langle \overline{w^{\prime\prime}\theta^{\prime\prime}}\right\rangle + \left\langle \overline{w^{\prime}\theta^{\prime}}\right\rangle\right) $ and $\left\langle\overline{w^{\prime}\theta^{\prime}}\right\rangle /\left(\left\langle \overline{w^{\prime\prime}\theta^{\prime\prime}}\right\rangle + \left\langle \overline{w^{\prime}\theta^{\prime}}\right\rangle\right)$ in (b) for all the cases with $\lambda_p=0.12$ and for increasing $\lambda_f$ (Slender06 to Wide32).
\begin{figure}
	\centerline{\includegraphics[scale=0.8]{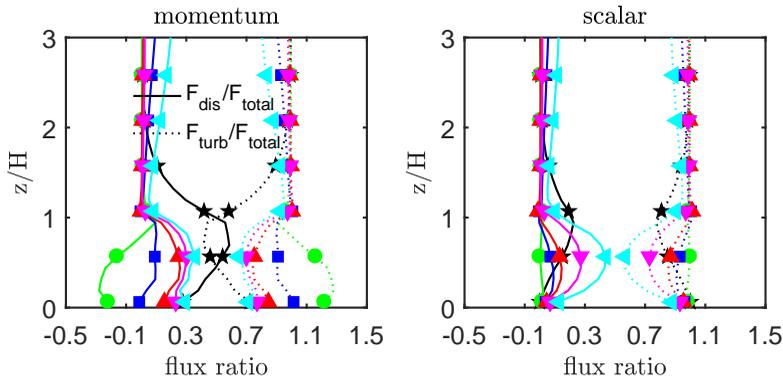}}%{Fig4_updated}}% Images in 100% size
	\caption{Fractions of $x-y$ averaged dispersive or turbulent (resolved $+$ subgrid-scale) fluxes for (a) momentum and (b) scalar.
		$\bigstar$,
		Slender06; $\circ$, Sf08; $\square$, Sf12; $\bigtriangleup$, Sf16; $\bigtriangledown$, Sf24; $\triangleleft$, Wide32.}
	\label{fig:disturb_ratio}
\end{figure}
Examples of previous studies on the momentum dispersive fluxes \citep{poggi2004note,christen20048,Coceal:2007fdba,Martilli:2007kbba,poggi2008effect,Leonardi:2015drba,Giometto:2016exba} and a few on dispersive scalar fluxes \citep{christen20048,Leonardi:2015drba} have demonstrated their importance within the canopy sublayer, as well as in the roughness sublayer. Similar to previous findings, our simulations also indicate that the dispersive fluxes are significant within the roughness arrays. The dispersive flux can contribute $\approx$ 50$\%$ of the total momentum or scalar flux (where the contribution of the sub-grid scale fluxes is included in the total) below the roughness elements height for some cases. Interestingly, there are no monotonous trends of how the dispersive fluxes vary with $\lambda_f$. The fractions of dispersive fluxes, in general, are the highest for the most eccentric geometries Slider06 and Wide32.

Note that \citet{Leonardi:2015drba} and \citet{Coceal:2007fdba} commented that the dispersive fluxes are only important on the intermediate time scales. We performed our analysis for the time averaged quantities over time spans of approximately $H/U_b$ = 1200 ($U_b$ is the bulk velocity) and longer than the $H/U_b$=600 suggested in \citet{Leonardi:2015drba}. Still, temporal averaging for large $H/U_b$ gives converging results of the dispersive fluxes that continue to show that they are significant for the canopy and roughness sublayers. Long-lived streamwise rolls can develop and persist in periodic numerical simulations due a ``locking" of these rolls as they begin to interact with themselves across the periodic boundaries. Such rolls would increase the estimated magnitude of the dispersive fluxes unless very long time averaging is performed. This might have been the argument of previous studies, but in our simulations the large dispersive fluxes are not a numerical artifact as will be further illustrated below.

Both dispersive stress and scalar flux are important portions of the total fluxes especially within the canopy sublayer, and their differences are now studied in detail. Figure \ref{fig:disp_uwsline} shows $u^{\prime\prime}_N =u^{\prime\prime}/\langle u(z=H)\rangle$ and $\theta^{\prime\prime} _N = \theta^{\prime\prime}/\Delta\theta$, where $\Delta\theta=\langle\theta(z=H)-\theta(z=0.125H)\rangle$ for an $x$-$z$ cross session indicated by the blue line in figure \ref{fig:setup}. 
The pseudocolor plots and streamlines along the horizontal line shown in figure \ref{fig:setup} are spatially averaged for all repeating units in the domain. Negative $u^{\prime\prime}_N$ in figure \ref{fig:disp_uwsline}(a) is due to loss of horizontal momentum $u$ in the wakes produced by the obstacle. While figure \ref{fig:disp_uwsline}(d) shows positive $\theta^{\prime\prime}_N$ as a result of the surfaces being kept at a higher temperature. As $\lambda_f$ increases, in figures \ref{fig:disp_uwsline}(b) and (e) and figures \ref{fig:disp_uwsline}(c) and (f), the most pronounced momentum-scale distinction can be found upstream of the obstacle where the mean recirculation pattern results in ``counter-gradient" dispersive momentum transport (fluid slowed by the pressure field generated by the obstacle being advected downwards) but ``down-gradient" scalar transport (cooler or lower concentration fluid transported downwards). This difference emerges from the non-local action of pressure on momentum (the fluid streamwise velocity has to decrease as it approaches the windward face even before it contacts that face), but its absence from the dynamics of scalars (the fluid has to ``touch" the surface to uptake scalars). The sign of the dispersive fluxes is thus the same as the total fluxes, while there is partial cancellation of dispersive fluxes for momentum. 

\begin{figure}
	\centerline{\includegraphics[scale=0.8]{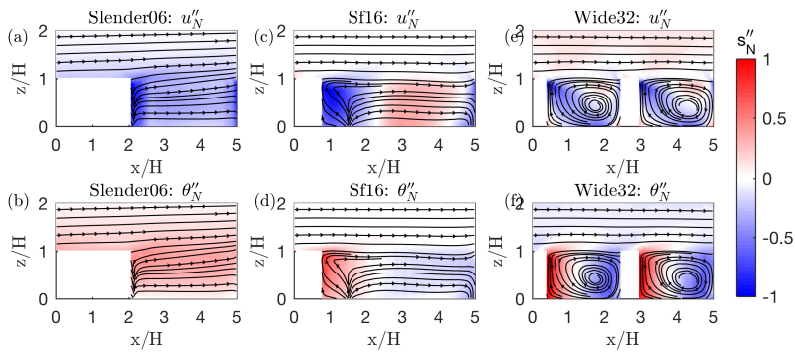}} %{Fig5}}% Images in 100% size
	\caption{Normalized dispersive stress and flux for three cases: color scale indicates $s^{\prime\prime}_N=s^{\prime\prime}/\Delta s $ for $s=u$ or $\theta$. Lines with arrows are the  $\overline{u}$ and $\overline{w}$ streamlines.
	}
	\label{fig:disp_uwsline}
\end{figure}

The quadrant analysis approach can also be applied to dispersive fluxes to compute a transport efficiency  $\eta=F_{total}/\langle F_{ejection}+F_{sweep}\rangle $ for momentum ($\eta^d_m$) and scalars ($\eta^d_\theta$). For example, when $u^{\prime\prime}<0$ and $w^{\prime\prime}>0$, we denote it as a dispersive sweep (this is no longer an event but a persistent spatial feature). Figures \ref{fig:disp_quadrant}a-c indicate that there is distinct dissimilarity between the vertical distribution of $\eta^d_\theta$ and $\eta^d_m$, related to the physical differences that arise from the role of pressure discussed above. In general, it is observed that $\eta_m^d < \eta_s^d$, except for case Slender06 shown in figure \ref{fig:disp_quadrant}d. This is in stark contrast to the turbulent transport efficiencies shown in figure \ref{fig:efficiency} where the slender case was the one showing a lower turbulent momentum transport efficiency. As $\lambda_f$ increases, a non-monotonic trend is observed for the difference between ``efficiency'' of dispersive transport of momentum and passive scalar, with the most eccentric geometries again showing the highest transport efficiencies. This is in agreement with the observation above that these eccentric cases also had the highest fraction of the fluxes carried by the dispersive part. 

\begin{figure}
	\centerline{\includegraphics[scale=0.75]{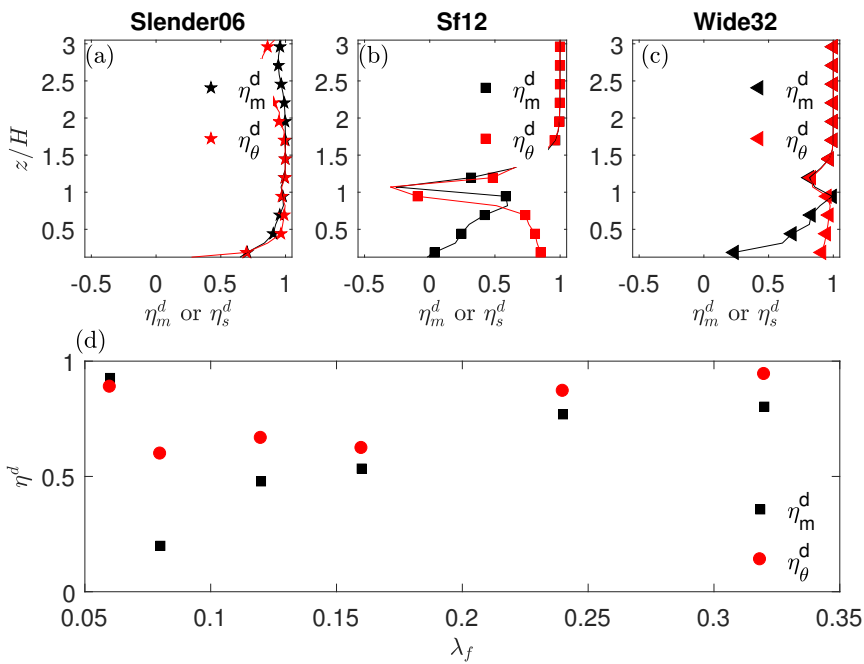}}   %{Fig5plus}}% Images in 100% size
	\caption{Dispersive transport efficiencies for different cases: 
		(a) $\bigstar$,
		Slender06; (b) $\square$, SF1225; (c) $\bigtriangleup$, Wide32; (d) $\eta_m^d$ and $\eta_\theta^d$ averaged for $z=0-2H$. }
	\label{fig:disp_quadrant}
\end{figure}

\section{Conclusion}

We use LES to compare and contrast the transport of momentum and passive scalars over \textit{very rough} surfaces consisting of three-dimensional cuboid roughness elements. The paper focuses on the comparisons between momentum and scalar exchanges, and between turbulent and dispersive contributions, over rough surfaces in the canopy and roughness sublayers. The effect of the frontal blockage of the roughness geometry on the fluxes is also investigated.

The influence of different three-dimensional roughness elements can be seen to alter the general turbulent flow characteristics. We observe a general transitioning behaviour from a surface-layer-like to a typical mixing-layer-like flow or `obstructed shear flow' when the frontal area ratio increases; this ratio $\lambda_f$ encodes the flow blockage by the roughness elements (the other widely studied geometric parameter, the planar density of the roughness $\lambda_p$, is maintained constant in our study). This is particularly illustrated by the decrease in the shear length scale and the dampened penetration of large eddies into the canopy layer as $\lambda_f$ increases. This transition causes a monotonic and significant increase in the momentum transport efficiency as the flow becomes more mixing-layer-like, while on the contrary these transport efficiencies for scalars measured by the correlation coefficient change more mildly and non-monotonically with increasing $\lambda_f$. This is mainly a result of the distinct contributions from large-scale motions in the roughness sublayer, which are able to increase the interaction between the canopy sublayer and the roughness sublayer more effectively for momentum than for scalars as the geometry is modified to increase blockage ratio.  

Turbulent transports of momentum and passive scalar are found to be similar in general, as evidenced by a quadrant analysis, with higher averaged turbulent momentum transfer efficiency. The unique exception is case Slender06, where the turbulent momentum and scalar transports difference becomes more substantial. However, the dispersive momentum and scalar fluxes show more pronounced differences than their turbulent counterparts (Slender06 again being the exception and displaying the strongest scalar-momentum similarity for the dispersive part). These dispersive fluxes are significant, particularly for the most eccentric geometries (Slender06 and Wide32) where they amount to $\approx 50\%$ of the total fluxes. The least eccentric geometries on the other hand show the most substantial discrepancy between scalar and momentum exchange. The differences between momentum and scalar dispersive fluxes are traced back to the non-local action of pressure in the momentum dispersive contributions, and the absence of a pressure-counterpart for scalar transport. This results in stronger interactions between roughness elements wakes as the surface geometry changes, giving rise to the characteristic recirculation patterns and leading to consistent differences between the dispersive transport efficiencies of momentum and scalar.

Although we only simulated flows with constant scalar boundary conditions, the results about dis(similarity) between turbulent and dispersive fluxes and between momentum and scalars should hold regardless of the scalar boundary condition. On the other hand, if the scalar concentration influences buoyancy (e.g. with large temperature fluctuations), the active role of the scalars will then have a strong impact on the dynamics \citep{Bou-Zeid:2011hrba} and the present results will be altered. This study demonstrates that the (dis)similarity between momentum and scalar transport in three-dimensional \textit{very rough} surfaces can be complicated by the the spatial variability of the roughness elements and the specific topology of the underlying geometry. However, the present findings, especially the significance of the scalar dispersive flux contributions over dense canopy, can inform the interpretation of experimental measurements, which are often point-wise data where dispersive fluxes cannot be estimated: such measurement in the canopy and roughness sublayers are missing nearly half of the total fluxes. In addition, in the dense canopy sublayer, the mechanistic difference (arising from the pressure term) in generating dispersive momentum and scalar fluxes need to be incorporated in model development. Current parameterizations either neglect dispersive fluxes or do not distinguish them from turbulent fluxes, but as our analyses show their physics are quite different and this difference is not the same for momentum and scalars.\\
 
\section{Acknowledgment}

This work was supported by the U.S. National Science Foundation's Sustainability Research
Network Cooperative Agreement 1444758 and Grant 1664091. The simulations were performed on the
supercomputing clusters of the National Center for Atmospheric Research through projects P36861020 and UPRI0007.

\appendix
\counterwithin{figure}{section}
\section{Resolution test}
A sensitivity test is carried out to examine the effect of the  number of points used in representing the obstacles. It was found previously by \cite{Tseng:2006fgba} that at least 6 points per dimension of the obstacle are required to represent a solid in LES (they used a similar approach and LES code to the ones we are using here). Therefore, we doubled the resolutions for the case Wide32 listed in table \ref{table1}. 
In addition, a test case with all other parameters being the same as Wide32 except that $L_x/\delta$ is doubled to 6.25 is also conducted. Figure \ref{fig:profiles_compare} shows the comparison between the original simulation, the higher resolution case and the case with longer domain length, where $\theta_1$ is $\langle\theta \rangle$ at $z=\delta$ and $\theta_0$ is $\langle\theta \rangle$ at $z=1/8H$. Both test cases were run for about 30 eddy turn over times and are averaged for the last 15 eddy turn over times. We compute the height-averaged relative mean-square-error defined as $r_e^n=\bigg \langle\frac{\sqrt{(q_n-q_0)^2}}{q_0}\bigg\rangle _z$, where $q_n$ is some quantity for comparison in case $n$ ($n=1$ being doubled $L_x$ and $n=2$ being the case of double resolution in all directions) and $q_0$ is that same quantity from the initial setup of case Wide32. $r_e^1$ in percentage for figures (a)-(d) are 1.7, 1.5, 15 and 27 $\%$; $r_e^2$ in percentage for figures (a)-(d) are 1.7, 2.0, 5.5 and 4.2 $\%$. $r_e^2$ are all within 6 $\%$ compared to case Wide32, which shows that we achieve good grid convergence. The larger deviation seen in case of doubled $L_x$ in the second-order moments is due to the fact that in the larger domain the streamwise roll are less persistent and the convergence to a linearly decreasing stress (as expected) is faster. However, for $z/H <2$, the deviation in the case of doubled $L_x$ is small, likely because of the dominance of the small-scale wall-attached eddies.
\begin{figure}
	\centerline{\includegraphics[scale=0.6]{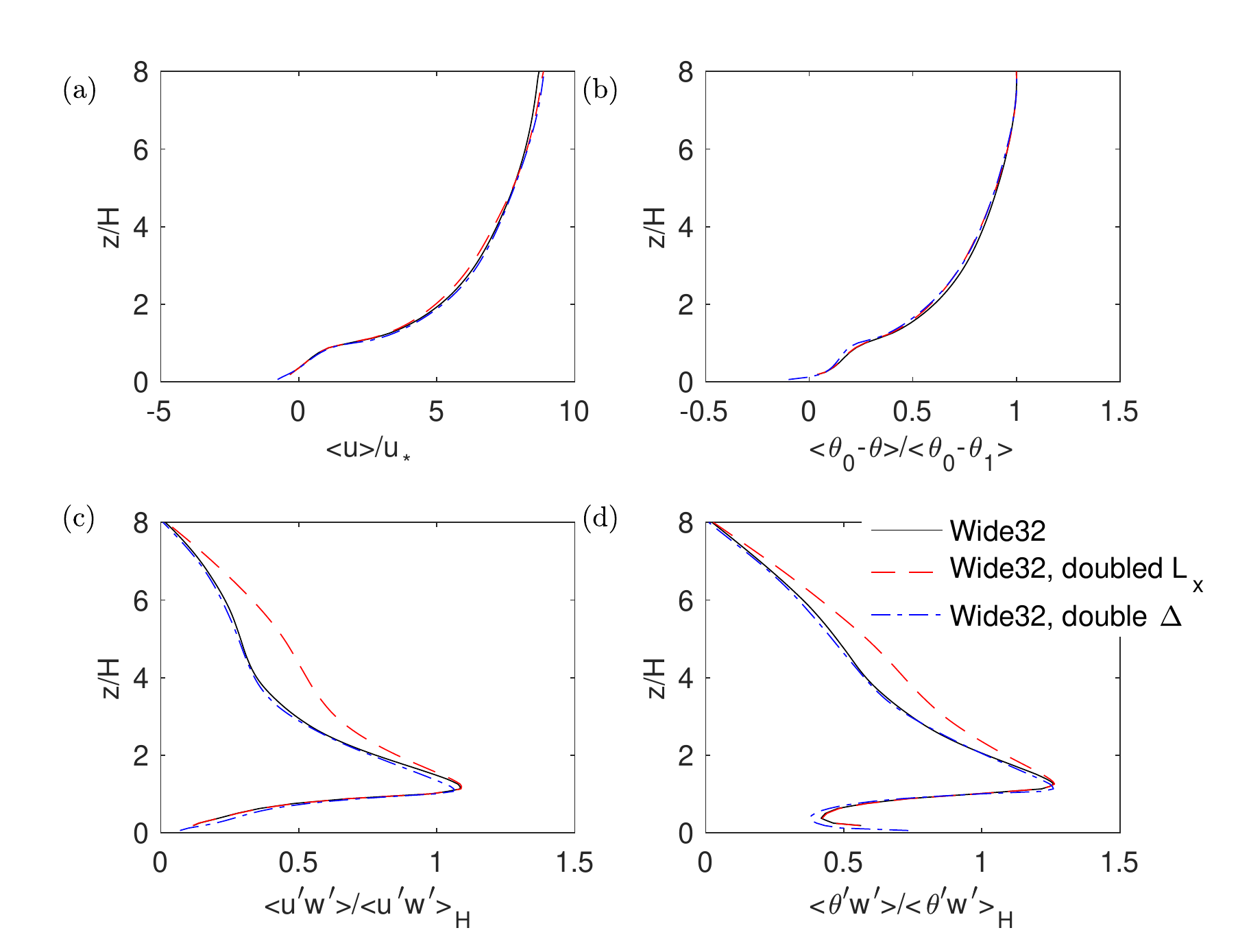}}% Images in 100% size
	\caption{Comparisons of horizontally averaged quantities. (a) $\langle u /u_*\rangle $, (b) $\langle (\theta_0 -\theta)/(\theta_0-\theta_1) \rangle$, where $\theta_0=\theta(z=0.125H)$ and $\theta_1=\theta(z=\delta)$ (c) $\langle u^\prime w^\prime /u_*^2 \rangle $(d) $\langle \theta^\prime w^\prime /u_*\theta_1 \rangle$.}
       
	\label{fig:profiles_compare}
\end{figure}

Figure \ref{fig:dispersive_compare} shows the percentage of dispersive momentum and scalar flux to total flux for these three cases. Good agreements between the three cases confirm that the differences observed between dispersive momentum and scalar fluxes are robust, especially for the roughness sublayer, which is the focus of this study. 

\begin{figure}
	\centering{\includegraphics[scale=0.6]{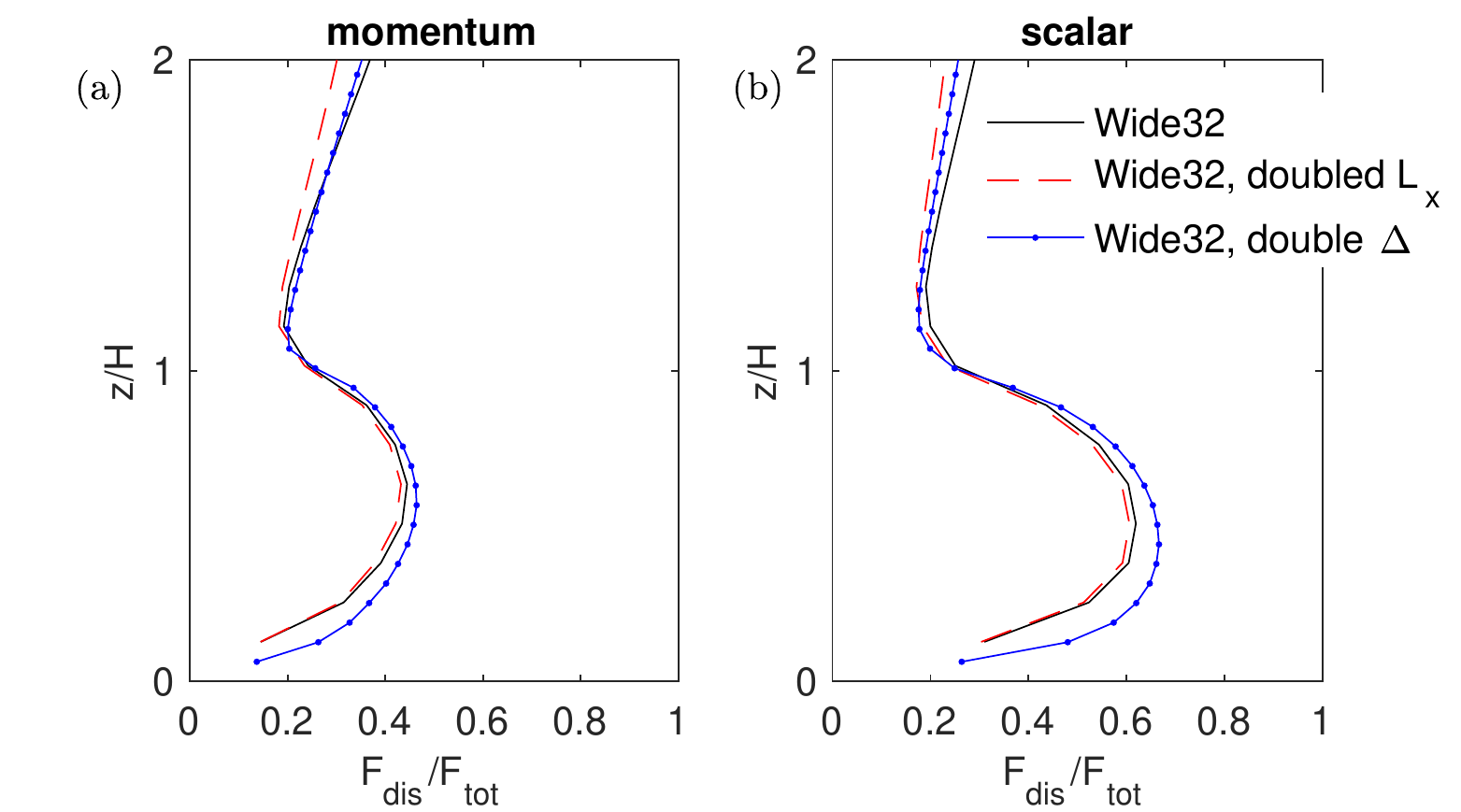}}% Images in 100% size
	\caption{Comparison of percentage dispersive fluxes. (a) momentum, (b) scalar.}
	\label{fig:dispersive_compare}
\end{figure}

\section{Quadrant analysis at different points}

\counterwithin{figure}{section}
Quadrant analysis similar to results in figure \ref{fig:qcontribution} are presented here for Slender06 in figures \ref{fig:quadrant_allpoints_momentum} and \ref{fig:quadrant_allpoints_scalar}; Wide32 in figures \ref{fig:quadrant_allpoints_momentum2} and \ref{fig:quadrant_allpoints_scalar2}.
\begin{figure}
	\centering{\includegraphics[scale=0.6]{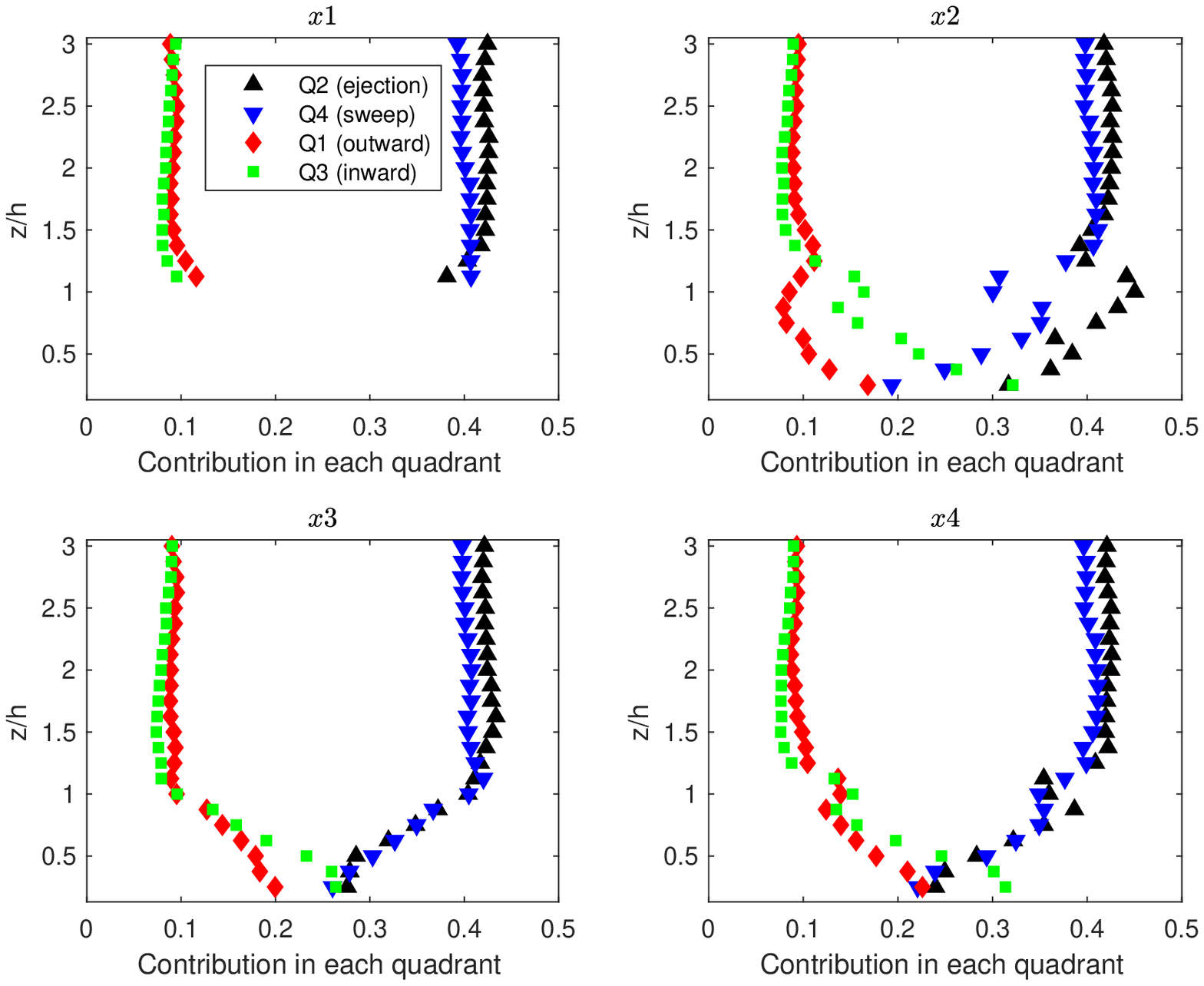}}% Images in 100% size
	\caption{Contribution to total flux for momentum for Slender06. (a)-(d): points $x1$-$x4$ indicated in figure \ref{fig:setup}. 
		Ejections are black $\bigtriangleup$
		; sweep are blue $\bigtriangledown$; outward interaction are red $\diamond$; and 
		inward interaction are green $\square$.}
	\label{fig:quadrant_allpoints_momentum}
\end{figure}

\begin{figure}
	\centering{\includegraphics[scale=0.6]{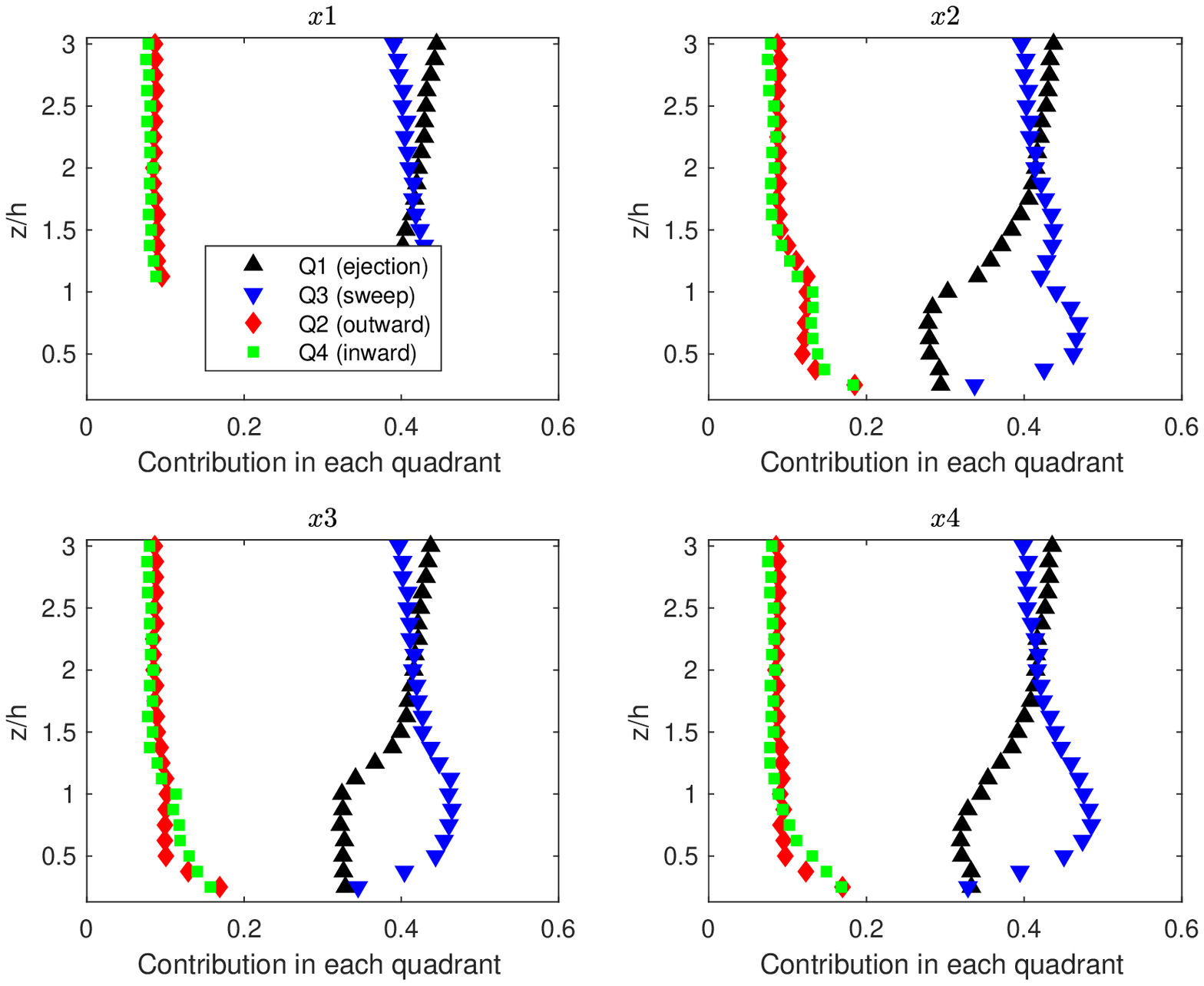}}% Images in 100% size
	\caption{Contribution to total flux for scalar for Slender06 (a)-(d): points $x1$-$x4$ indicated in figure \ref{fig:setup}. 
		Ejections are black $\bigtriangleup$
		; sweep are blue $\bigtriangledown$; outward interaction are red $\diamond$; and 
		inward interaction are green $\square$.}
	\label{fig:quadrant_allpoints_scalar}
\end{figure}

\begin{figure}
	\centering{\includegraphics[scale=0.6]{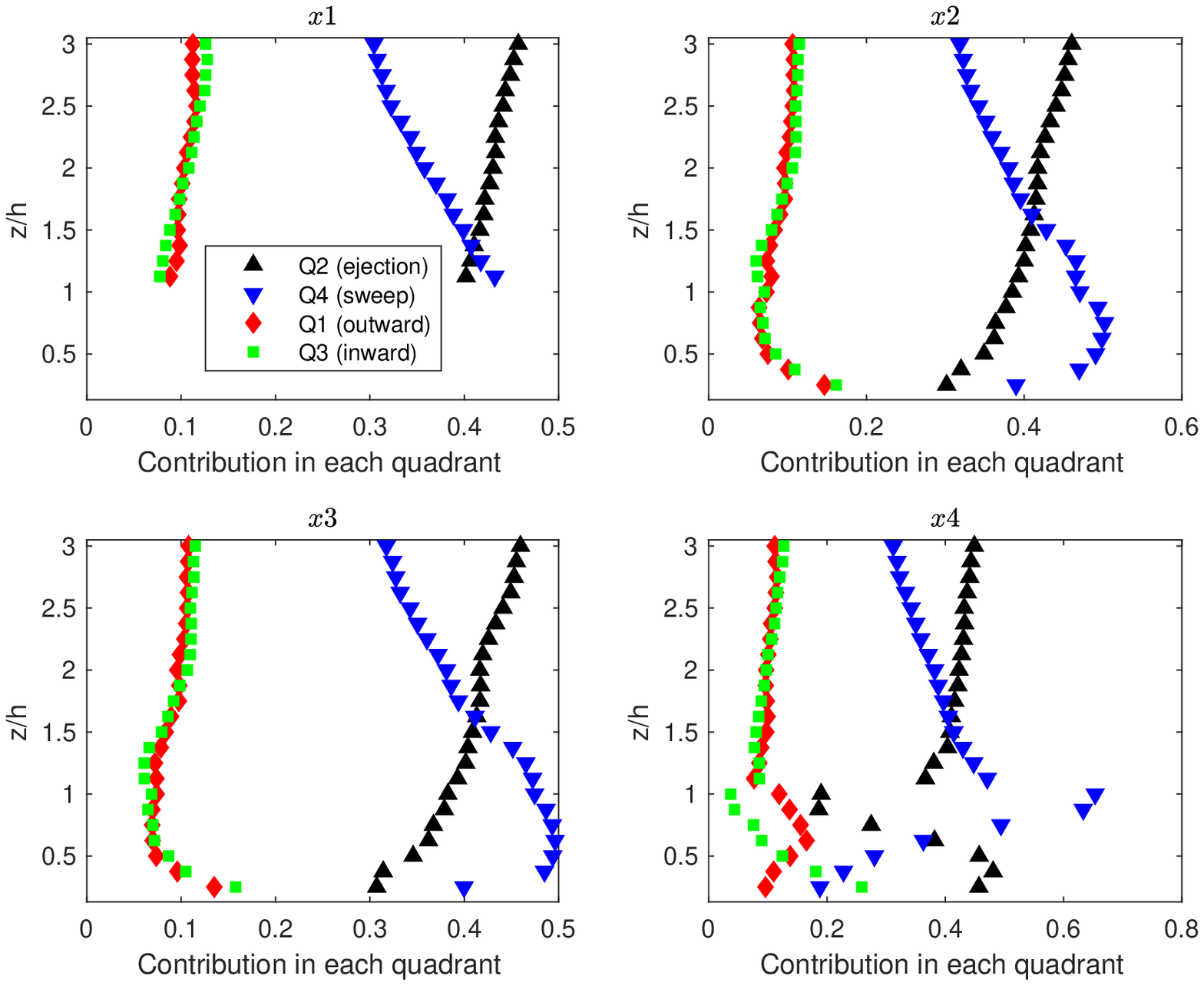}}% Images in 100% size
	\caption{Contribution to total flux for momentum for Wide32 (a)-(d): points $x1$-$x4$ indicated in figure \ref{fig:setup}. 
		Ejections are black $\bigtriangleup$
		; sweep are blue $\bigtriangledown$; outward interaction are red $\diamond$; and 
		inward interaction are green $\square$.}
	\label{fig:quadrant_allpoints_momentum2}
\end{figure}

\begin{figure}
	\centering{\includegraphics[scale=0.6]{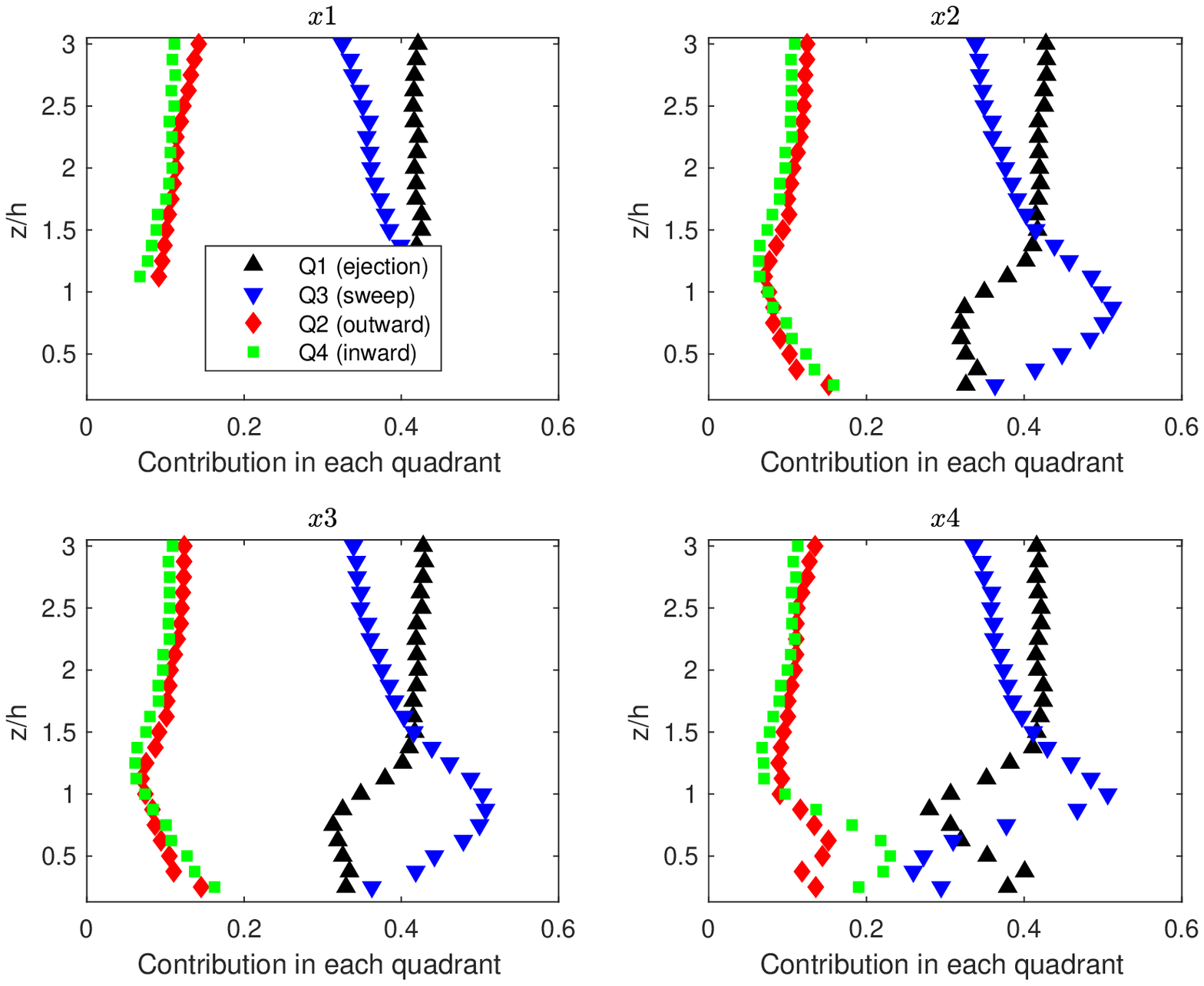}}% Images in 100% size
	\caption{Contribution to total flux for scalar for Wide32 (a)-(d): points $x1$-$x4$ indicated in figure \ref{fig:setup}. 
		Ejections are black $\bigtriangleup$
		; sweep are blue $\bigtriangledown$; outward interaction are red $\diamond$; and 
		inward interaction are green $\square$.}
	\label{fig:quadrant_allpoints_scalar2}
\end{figure}

\newpage
\bibliographystyle{jfm}
\bibliography{reference}

\end{document}